# Three consequences of decompositional consistency

Chris Fields

21 Rue des Lavandières
Caunes Minervois 11160 France

chris@hayfields-communications.com

**Abstract**

Decompositions of the world into systems have typically been regarded as arbitrary extra-theoretical assumptions in discussions of quantum measurement. One can instead regard decompositions as part of the theory, and ask what conditions they must satisfy for self-consistency. It is shown that self-consistent decompositions that specify a measurement context (i.e. von Neumann decompositions) must represent apparatus as containing internal decohering environments and as having known pointer components and ready states. Under these circumstances a von Neumann decomposition can function as a component of the ready state of the observer. Minimal no-collapse quantum mechanics supplemented by these consistency requirements on von Neumann decompositions avoids the measurement problem.

## 1. Introduction

At the end of his now-classic "rough guide" to decoherence, W. Zurek [1] points out the critical role in the analysis of measurement that is played by the decomposition of the universe into interacting systems, and remarks that a principled understanding of the origin of such decompositions would be "most useful" (p. 1818). It is the indeed the universal practice of physicists to choose whatever decompositions seem to offer advantage, and of human beings in general to draw conceptual boundaries around bits of the world, give them names, and consider them as entities, all with apparently complete freedom. Such decompositions are, however, or if viewed as abstractions are at least *represented by*, physical states of those who employ them. Hence they cannot be arbitrary: every instance of a decomposition is a projection, into some coordinate system that captures an observer's cognitive activity, of the universal state vector $\Psi_U$, and as such must be producible by some physical process or other. Zurek's remark on the utility of understanding the origin of decompositions thus points, at least in part, to the evident utility of understanding the origins of and constraints on physical implementations of cognition.

The present paper approaches Zurek's question indirectly by focusing on a particular class of decompositions, the von Neumann decompositions that form the basis for the analyses of measurement interactions. These decompositions divide the world into a system of interest (**S**), the states of which are unknown and are the target of the measurement, an apparatus (**A**), and embedding environment (**E**), and an observer (**O**). They represent the context of the measurement interaction from **O**'s point of view, and can be viewed as encoding **O**'s reportable knowledge of the both the context and the interaction. This paper asks what information such a von Neumann decomposition {**S,A,E,O**} must contain to both adequately and consistently support the analysis of a measurement interaction. It first shows that von Neumann decompositions that represent **A** as linearly coupled to **S** are not self-consistent, and that removing this inconsistency causes the measurement problem, as standardly conceived from the observer's perspective, to disappear. Second, it shows that in the context of a consistent von Neumann decomposition, einselection-based stability criteria that identify non-destructively observable states [1, 2] are insufficient to identify pointer bases. The attempt to recover the macroscopic world from einselection alone – the existential interpretation – therefore fails: einselection in effect yields a classical "scene" but no objects. Finally, it shows that the information contained in a consistent von Neumann decomposition is the contextual information required to define the "ready" state of an observer. This dynamic interpretation of the decomposition allows it to be represented as a data structure with a single argument – the specification of **A** – and reframes Zurek's question in the special case of human observers as an empirical question that is not only addressable, but at this stage significantly answerable by neuroscience. The paper closes by examining some implications of these three consequences of decompositional consistency for the interpretation of measurement, and by arguing in particular that minimal no-collapse quantum mechanics requires only self-consistency of von Neumann decompositions to avoid the measurement problem.

Before proceeding, it is worth reviewing some of the intuitions that go into a von Neumann decomposition {**S,A,E,O**} and making them at least slightly more precise. First, the environment **E** is included in the decomposition to indicate explicitly that the universe lacks *a priori* boundaries that physically isolate quantum systems of interest; all real systems are open to some environment [1-6]. The environment **E** in a decomposition {**S,A,E,O**} comprises all degrees of freedom not explicitly allocated to **S**, **A** or **O** and is assumed to be large compared to any of these systems. Moreover, the only states of **E** that are regarded as relevant to the measurement are those that linearly encode pointer values of **A**; all other environmental states are ignored. Second, as noted above, a decomposition employed by an observer **O** to define the context of a measurement interaction must be represented by a physical state of **O**; this representation is what allows **O** to report what was measured as well as the result. Hence a minimal condition on **O** is that its physical structure includes degrees of freedom sufficient to represent both the decomposition itself and the results obtained from measurements employing the decomposition. The plural "results" is used here advisedly; a "measurement" that can be made once but never replicated may have poetic value, but is of little use in science. This implies a minimal condition on **A**: the physical structure of **A** must be such as to allow its re-identification and hence re-use over time by **O**. These conditions can be summarized by saying that **E** transmits but does not transform information, **O** represents the decomposition as well as the results, and **A** cannot be an unknown quantum state.

Besides these conditions on its elements, a von Neumann decomposition involves three assumptions about couplings. First, **S** is coupled to **A**, otherwise **A** would be irrelevant to the decomposition. Second, the **S**-**E** and **S**-**O** interactions are negligible, otherwise **S** would be decohered by **E** before the measurement could be made, or **O** could obtain information about **S** directly without need of **A**. Finally, **A** and **O** are coupled to **E** but not to each other, otherwise **E** would be irrelevant. These coupling conditions can be summarized by saying that observers use apparatus to make measurements, they interact with their apparatus via ambient environmental media, and measurement only works if the apparatus being employed actually interacts with the system of interest.

## 2. Decompositional consistency eliminates the observer's measurement problem

Following the analysis of Tegmark [3], it has become widely if not universally accepted that internal representational states – e.g. memories – in human brains, and by extension, equivalent internal structures of any other candidate observers, are fully decohered by the brain's internal environment [2, 4 – 6]. While Tegmark proposes particular basis states – the resting versus firing states of action-potential generating neurons – as underlying internal cognitive representations, his analysis succeeds for any proposed decomposition of the brain in which the degrees of freedom to which an internal representational role is assigned interact sufficiently strongly with a sufficiently larger collection of degrees of freedom to which an internal representational role is not assigned. Tegmark does not explicitly consider this case, but it is clear that his analysis applies not only to internal representational states directly consequent to perceptual input, but also to those that originate by internal processes, for example states representing novel theoretical ideas. While the details of the decoherence times estimated in Tegmark's analysis may be questioned [7, 8], one significant consequence is clear: the external output states of human brains, and again by extension the equivalent external structures of any other candidate observers, are causal consequences of fully decohered internal states and are therefore of determinate content. Thus even if one is inclined to argue that some internal cognitive processes briefly employ irreducibly quantum computation, there is no question but that outputs overtly reported by human observers, including outputs of purely internal cognitive processes, are fully decohered and thus effectively classical.

As Tegmark [3] points out, internal decoherence is directly relevant to quantum computation. It is the imbalance in the roles attributed to component degrees of freedom that makes decoherence such a problem for quantum computing: only a few degrees of freedom in any physically-realized quantum computer are interpreted as information-bearing qubits; the rest compose a potentially-decohering internal environment. This decohering internal environment cannot be dispensed with, as it is essential to the functioning of the device; it particular, it is essential to the process of measuring, and hence outputting, the data represented by the qubits. Thus even though some internal processes are in this case explicitly acknowledged to involve irreducibly quantum computation, there is no question but that outputs overtly reported by any physically-realized quantum computer are decohered and thus effectively classical.

The basic structure of Tegmark's [3] analysis of observers, and of quantum computers as a class of apparatus, would appear to apply to *any* physically-realizable apparatus; Tegmark's analysis would appear to imply that any physically-realizable apparatus would incorporate an internal decohering environment, that the internal decohering environment would efficiently decohere any apparatus states that became entangled with quantum systems, and that any apparatus would therefore present a fully-decohered pointer state to the external environment. However, prominent recent reviews of the role of decoherence in quantum measurement, all of which cite Tegmark's analysis with approval, continue to represent the pointer state of the apparatus prior to interaction with the external environment **E** by a linear superposition $\Sigma_i \lambda_i |s_i\rangle|a_i\rangle$, where the $|s_i\rangle$ are basis vectors for **S** and the $|a_i\rangle$ are the pointer basis for **A** [4 – 6]. Such a linear representation, assuming more than one of the $\lambda_i$ are non-zero, implies that the pointer state of **A** has not yet been decohered, i.e. that **A** does not incorporate a decohering internal environment. This is confirmed by discussions of decoherence and einselection being imposed on **A** by the *external* environment, the environment **E** via which pointer states of **A** are observed by **O**. While this approach accords with the pedagogical tradition of treating quantum systems first as closed and only later as open to **E**, Tegmark's analysis raises the question of whether a linear representation of **A** is consistent with **A** being viewed as a physically-realized apparatus that is employed in a measurement context. It is therefore worth asking whether a linear representation of **A** is consistent with the assumptions and constraints of a von Neumann decomposition.

Let us first consider the case ("Case 1") in which **A** incorporates an internal decohering environment with which **O** does not interact. In this case, **S** interacts with internal components of **A**, producing an entangled state $|S\rangle|A^{internal}\rangle$. This correlated state is decohered by the internal environment of **A**, which can be assumed without loss of generality to destroy $|S\rangle$. The resulting decohered state, which can be represented $|A^{d(internal)}\rangle$, is correlated by **A**'s internal dynamics with a single pointer state $|a_n\rangle$. These internal dynamics are executed by the internal environment of **A**, but unlike that of the external environment **E**, the representation of the internal environment of **A** in a von Neumann decomposition is not restricted to the transfer of information only; thus **A**'s internal dynamics can be represented as non-trivial (i.e. energy dissipating and non-linear), as indeed they are in the case of actual apparatus. Provided that the resulting pointer state $|a_n\rangle$ commutes with the **A**-**E** interaction, it is encoded by a determinate environmental state $|e_n\rangle$. Up to considerations of classical noise, this state $|e_n\rangle$ transmits information only. The observer "samples" $|e_n\rangle$ in the sense described by Zurek [1], i.e. interacts with $|e_n\rangle$ in a way that is effectively nondestructive. Assuming that **O** associates $|e_n\rangle$ with $|a_n\rangle$ and that **O**'s internal representations are fully decohered by **O**'s internal environment, **O** is able to remember or externally report $|a_n\rangle$. This scenario is precisely that envisioned by designers of quantum computers and measurement apparatus, and is the most natural interpretation of experiments involving not only micro- but also meso- and macroscale quantum states [6, 9]. It can be consistently modeled by minimal no-collapse quantum mechanics, as it involves only virtual collapse – information loss to the internal environment – via internal decohering interactions. The observer **O** is external to **A**, does not interact with **A**'s internal environment, and so is never faced with an environmental encoding of a superposition; hence from **O**'s perspective the measurement problem as traditionally conceived does not arise.

The second, more traditional characterization of **A** involves two sub-cases, one in which **A** incorporates a decohering environment with which **O** does interact, and the other in which **A** incorporates no internal decohering environment. The first of these sub-cases ("Case 2a") is equivalent to a situation in which **A** is inter-penetrated by the external environment **E**; **O**'s access to the internal environment renders it continuous with **E**. Tegmark [3] explicitly employs a decomposition of this type, confining **A** to "the pointer position on the measuring apparatus" and relegating the remainder of the physical realization of the apparatus to the environment (p. 4196). In this case, both the pointer states $|a_i\rangle$ and the system state $|S\rangle$ are directly coupled only to **E**. But **E** is large, so both the $|a_i\rangle$ and $|S\rangle$ will be immediately decohered. Because the $|a_i\rangle$ and $|S\rangle$ are not directly coupled, the correlation $\Sigma_i \lambda_i |s_i\rangle|a_i\rangle$ is never established and the decohered pointer state $|a_n\rangle$ conveys no information about $|S\rangle$. Hence while this representation of **A** is adequate to discuss interactions between **A** and **O** – the purpose to which Tegmark puts it – it is inadequate to discuss **O**'s use of **A** to obtain information about **S**. The measurement problem does not arise in this case because there is no measurement: no information about the state of **S** is available to **O**. The intuitive notion of a von Neumann decomposition disallows this case by requiring that the **S**-**E** interaction be negligible for precisely this reason.

The second traditional sub-case ("Case 2b") involves an apparatus **A** that incorporates no internal decohering environment. *All* degrees of freedom of **A** are therefore coupled directly to **S**; if any degrees of freedom of **A** existed that were not coupled directly to **S**, they would constitute an internal decohering environment, contrary to assumption. One can view this situation in one of two ways. If decoherence by the external environment is temporarily neglected, the correlated state $\Sigma_i \lambda_i |s_i\rangle|a_i\rangle$ is an isolated, unknown (by assumption, as **S** is assumed to be unknown) quantum state. By the no-cloning theorem [10], it cannot be replicated; hence **O** cannot re-identify it by observation. This is the situation envisioned in considerations of fully-linear measurement prior to the recognition that all real quantum systems are open; **O** was imagined to be directly coupled to this unknown quantum state, and hence to exist is a non-re-identifiable state $\Sigma_i \lambda_i |o_i\rangle$, giving rise to the measurement problem. With the recognition that **O** is only directly coupled to **E**, this view is no longer tenable. Decoherence by **E** must be taken into account, in which case since **E** is large, the correlated state $\psi = \Sigma_i \lambda_i |s_i\rangle|a_i\rangle$ must be regarded as immediately decohered. The environmental encoding of this state, however, is problematic; the "pointer basis" that is encoded will be determined not by the **E**-**A** interaction $H_{EA}$ alone, but by the composition of $H_{EA}$ and the **E**-**S** interaction $H_{ES}$. The latter interaction cannot be neglected, because it involves at least the degrees of freedom of the former. There is no reason to assume that $H_{AE}$ and $H_{ES}$ commute; indeed $H_{ES}$ is in general unknown. The observer is thus faced with an unknown and unknowable encoding state $H_{EA}H_{ES}(\psi) + H_{ES}H_{EA}(\psi)$, where there is no guarantee and no reason to assume that $H_{EA}H_{ES}$ and $H_{ES}H_{EA}$ share eigenstates. Such an encoding does not represent the pointer state of an *apparatus*; it has effectively replaced the original decomposition {**S**,**A**,**E**,**O**} with the ambiguous {**SA**,**E**,**O**}. An observer with this latter decomposition can only be regarded as not knowing – i.e. not being able to report – what she is interacting with. This is not a coherent description of measurement. The problem here is clear: Case 2b decompositions, in which **A** is represented as having no internal decohering environment, are effectively treating **S** as both open (exposed to the external environment) and closed (linearly coupled to the |$a_i$>). They are not self-consistent.

Because it is so facile to represent a Case 2b decomposition formally as a von Neumann chain $\Sigma_i \lambda_i$ |$s_i$> → $\Sigma_i \lambda_i$ |$s_i$>|$a_i$> → $\Sigma_i \lambda_i$ |$s_i$>|$a_i$>|$e_i$> → $\Sigma_i \lambda_i$ |$s_i$>|$a_i$>|$e_i$>|$o_i$>, it is worth reflecting on what this case entails in a context of a laboratory setting. Suppose **S** is an energetic proton-proton collision, and **O** observes the collision via a detector array connected eventually to a computer screen. In a Case 2b decomposition, everything except the multi-particle state immediately following the collision and the pixel excitations on the screen is considered part of the environment **E**, and **E** is restricted to the linear transmission, without modification, of information. Hence the decomposition can be written {**S**,{|$p_i$>},**E**,**O**}, where the |$p_i$> are the possible pixel excitation states. This decomposition is represented by a physical state of **O**, and constitutes **O**'s knowledge of the experiment. With this information, which includes none of what would ordinarily be considered a description of the detector and computer apparatus, **O** is expected to cope with an environmental encoding generated by the composition of the interactions between ambient photons and pixels on the one hand, and between an explosion of energetic particles and the rest of the local environment on the other. These interactions are not separated by a decohering environment, but are directly coupled. Without knowing what the apparatus is – without knowing more than just the pointer states – and representing it as having some internal dynamics, **O** simply cannot do this. With a Case 2b decomposition, **O** is faced with far worse than the traditional problem of observing a superposition; she is faced with observing a superposition that she does not have the information to specify. The von Neumann chain has broken down at step 3, because $H_{ES}$ cannot be ignored.

The three abstract cases and the Case 2b *gedankenexperiment* outlined above demonstrate two things. First, the measurement problem as traditionally conceived, i.e. the problem of an observer potentially faced with the experience of a quantum superposition, does not arise in von Neumann decompositions in which the apparatus is represented self-consistently. It *appears* to arise in decompositions in which **A** is regarded as not incorporating an internal decohering environment, but these decompositions are not self-consistent: they treat **S** as both open and closed. Second, it is clear from the discussion of Case 2 that the idea of an "apparatus" comprising only a collection of pointer states is incoherent. Such an "apparatus" cannot be coupled, in any self-consistent decomposition, to a system of interest **S** in a way that allows the acquisition and transmission of information about **S**; hence there is no sense in which the "pointer states" are states of an informative pointer. The first of these conclusions vindicates Bohr's insistence that apparatus be treated classically from within the decoherence program. The second provides a formal version of the experimentalist's intuition that an apparatus must be more than just its pointer.

There is, of course, a nagging concern here: the concern that the measurement problem has been closed up inside the apparatus with a simple dictum that because it no longer concerns **O**, it is no longer an issue. Bacciagaluppi [11], for example, formulates the measurement problem in a way that is independent of the "experience" of the observer. This larger issue will be considered below; for now, it is enough to note that the measurement problem does not arise *inside* **A**. Nothing inside **A** "experiences" a collapsing wave function; decohering interactions spread coherence information throughout the internal environment of **A**, but assuming that **A** dissipates sufficient energy, they will have negligible effect on its internal dynamics. Indeed, from the perspective of any given component of **A**, the local internal environmental encoding of residual coherence from **S** will be indistinguishable from the local internal environmental encoding of residual coherence from other components of **A**; i.e. it will carry no information about **S**. Such residual coherence will eventually leak out into the larger environment **E**. From this point on, it becomes part of the observer-independent measurement problem as formulated by Bacciagaluppi, and becomes subject to further decoherence processes within **E**.

### 3. Stability criteria are insufficient to identify pointer bases

The ability of observers to gather information about the world depends, in the existential interpretation [1, 2], on their ability to use stability criteria such as the predictability sieve to discover einselected pointer bases. Stability criteria identify states that are non-destructively observable (hereafter "NDO"); they approximate the commutivity criterion $[|a_i><a_i|, H_{AE}] = 0$ for defining states $|a_i>$ that are NDO by an **A**-**E** interaction $H_{AE}$. Zurek [1, 2] presents these criteria in the context of a Case 2 decomposition in which **A** is identified with its pointer states $|a_i>$; indeed, **A** is identified with the states that are found to be NDO by stability criteria, and which are then taken to be its pointer states. As shown above, however, Case 2 decompositions are either inadequate to describe measurement (Case 2a) or inconsistent (Case 2b). The question therefore naturally arises as to whether einselection is sufficient to identify pointer states in Case 1 decompositions, i.e. in the case of apparatus that are represented as having internal decohering environments with which **O** does not interact.

This sufficiency question is equivalent to the question of whether all NDO states of an apparatus are pointer states, which is in turn equivalent to the question of whether **A** can be consistently represented, in the context of a von Neumann decomposition, as bounded against **E** solely by its pointer states. A pointer state conveys information about the state of the measured system **S**; in the context of a Case 1 decomposition, it is an NDO state of **A** that is classically correlated with a determinate measured state of **S** by **A**'s internal dynamics, i.e. by its internal decohering environment. The sufficiency question can therefore be further reformulated as the question of whether **A** can be consistently represented, in the context of a Case 1 von Neumann decomposition, as bounded against **E** solely by states that carry information about the state of **S**.

Before proceeding formally, it is useful to consider an example. Apparatus used to interact with quantum systems dissipate energy. It is typically possible to turn them off. They therefore have an NDO state, typically the state of a switch or an indicator lamp, that encodes information about whether the apparatus is turned on, and hence whether it can be "ready" to make a measurement. An observer needs to know whether an apparatus is ready, therefore the state that indicates readiness is an important part of the specification of **A** within a decomposition. The state of the power indicator, however, does not in general encode any information about the state of **S**; it is relevant to the measurement outcome only as a gating condition. Hence it is an NDO state that is not a pointer state. For at least some apparatus, therefore, not all NDO states are pointer states, so stability criteria that identify NDO states are, at least for some apparatus, insufficient to identify pointer states.

This example can clearly be generalized. Any apparatus must have a state $|\mathbf{A}^{ready}\rangle$ such that the "pre-measurement" **S**-**A** interaction $H_{SA}$: $|\mathbf{S}\rangle|\mathbf{A}^{ready}\rangle \rightarrow |\mathbf{S}\rangle|\mathbf{A}^{internal}\rangle$. An observer must reportably know what $|\mathbf{A}^{ready}\rangle$ is, and must be able to determine non-destructively whether **A** is in this state. The ready state does not, however, encode the state of **S**; **A** is in $|\mathbf{A}^{ready}\rangle$ before the interaction. Therefore any **A** must be represented as having at least one NDO state, $|\mathbf{A}^{ready}\rangle$, that is not a pointer state.

The vast majority of NDO states of an apparatus must, however, be represented in a Case 1 decomposition as not encoding *any* available information, whether about **S** or about measurement-relevant states of **A**. Such states correspond to physical degrees of freedom that compose the boundary against **E** of the internal decohering environment of **A**. The number of degrees of freedom in the internal decohering environment must be large compared to the number of states encoding observer-accessible information about either **S** or **A**, as the internal environment must be represented as decohering states of both these kinds. The number of degrees of freedom exposed to **E** on the boundary of this internal environment scales as $N/l$, where N is the total number of degrees of freedom of **A** and $l$ is a representative dimension of **A**. The states corresponding to these boundary degrees of freedom must be NDO, because **A** must be re-identifiable over time by **O**. They cannot, however, be represented as informative, because to do so would disallow their role as unmonitored sinks for coherence information.

The uninformative NDO states required by a Case 1 representation of **A** correspond to the "uninteresting" states of the exterior of an apparatus, such as the states of its outer casing. These states must be accessible to observers for re-identifiability, but they are roundly ignored unless they change – unless, for example, someone has painted a big red "X" on an instrument's casing. A sudden change renders such states informative about **A**, not about **S**; they are not potential or redundant pointer states. They must be included in a Case 1 decompositional specification of **A** because to exclude them would render the internal environment of **A** unbounded against **E**, and hence undefined. But Case 1 decompositions are the only consistent decompositions adequate to the description of measurement, so all self-consistent apparatus specifications must include NDO states that are not pointer states; indeed the majority of NDO states in a self-consistent apparatus specification are not pointer states. Hence stability criteria that identify NDO states are insufficient to the identify pointer states, and therefore insufficient to identify the pointer basis.

The above result indicates that a self-consistent specification of an apparatus **A** in the context of a decomposition {**S,A,E,O**} in which the states of **S** are unknown must incorporate a specification of **A**'s pointer basis. The pointer basis cannot be discovered by observation unless the states of **S** are known independently, or a new decomposition is introduced that renders known internal states of **A** subject to manipulation. This reasoning clearly applies equally to |**A**$^{ready}$>, and to any other externally-accessible states that provide information relevant to measurement. The scope of the result is, however, larger than this. Let **A** represent any spatially bounded subset of the universe, and let **P** represent any proper component of **A** that has degrees of freedom on **A**'s boundary. Provided states of **P** do not exhaust the NDO states of **A**, **P** cannot be identified by stability criteria alone. Information obtained via alternative decompositions that allow access to internal states of **A**, or to known external systems that interact with **A** are always necessary. A key objective of the existential interpretation, the emergence of the appearance of a classical world by einselection alone, is therefore not feasible. Distinct classical objects cannot be obtained by non-destructive observation alone without an infinite regress of decompositions. This fundamental problem will be considered further below; for now it is enough to note that such an infinite regress cannot occur, so some knowledge not obtained through decomposition-mediated observation is required to produce the apparently classical world that we observe.

### 4. Decompositions function dynamically as ready states

The two consequences of decompositional consistency outlined above reveal that von Neumann decompositions must have an information-rich structure that is somewhat surprising initially, but on reflection seems rather obvious. A von Neumann decomposition describes a context in which a measurement is carried out by an observer using an apparatus. It represents the apparatus **A** explicitly and in considerable detail: it specifies an observationally re-identifiable spatial boundary, a set of pointer states, a ready state, and an internal dynamics that correlates the pointer states with the states of systems with which **A** interacts. In contrast, the decomposition specifies the system of interest **S** completely implicitly: **S** is whatever **A** is interpreted as interacting with, i.e. whatever is *in fact* correlated post-interaction with the observed pointer state. The decomposition gives **S** a name, but it is merely a placeholder for this uncharacterized, result-producing system. The decomposition specifies the environment **E** equally implicitly, as everything else bar the observer, with the sole caveat that states of **E** linearly encode boundary states of **A**, including the pointer states. Finally, it specifies the observer **O** as a re-identifiable system that represents the decomposition. The re-identifiability of **O** is a requirement of the decomposition's repeated utility; **O** or others must be able to compare results obtained by **O**, and hence must be able to re-identify **O** as a reporter of results.

Given this characterization, the dynamic role of von Neumann decompositions in measurement becomes clear. The observer, like the apparatus, must have a ready state: a state $|O^{ready}\rangle$ such that the **E**-**O** interaction $H_{EO}$: $|E\rangle|O^{ready}\rangle \rightarrow |E\rangle|O^{internal}\rangle$, where $|O^{internal}\rangle$ is processed by **O**'s internal dynamics into a reportable and rememberable result. The contents of three of the bits in $|O^{ready}\rangle$ are obvious: **O** must be "turned on" (i.e. aware), oriented toward the pointer of **A** (i.e. paying attention), and have sufficient available memory and processing capacity to store and process $|O^{internal}\rangle$ (i.e. be energetically capable of making an observation). The second of these bits of $|O^{ready}\rangle$ refers explicitly to a decomposition; therefore some of the information specified by a decomposition must be available in the ready state. A consideration of what a ready observer must be able to report in response to queries shows that *all* of the information in the von Neumann decomposition is in fact required in $|O^{ready}\rangle$. An observer who could not report the observable boundary of **A**, the identity of **A**'s pointer, and the facts that **A** was ready, that **A**'s dynamics were coupled to **S**, that the local environment faithfully transmitted information, and that she herself knew these things would not be "ready" to make a measurement. The internal dynamics of **O** must, moreover, have access to the information contained in the decomposition to compute inferences about the state of **S** from $|E\rangle$, even if **S** is only described as being in "the state correlated with $|a_n\rangle$" for some value $|a_n\rangle$ of **A**'s pointer. Hence the requirement that decompositions are represented by physical states of observers is not merely a philosophical consequence of a "materialist" metaphysical stance; it is essential to understanding the measurement process in physical terms.

It is now possible to advance a first part of an answer to Zurek's [1] closing question: a von Neumann decomposition is all but three bits of the ready state of an observer. It is a data structure in which, for a given observer, **O** is fixed, **S** and **E** are computed within the data structure, and **A** is a four-component free variable. In this representation, Zurek's question decomposes into two parts: 1) where does the data structure itself come from? and 2) where do the values of **A** come from? These are not foundational questions: they are straightforward empirical questions about any identifiable class of observers. These questions have yet to be fully answered in the case of human observers, but the basic contours of their answers are clear. The first question falls to comparative functional neuroscience, and is answered by models of multi-component binding in high-level perceptual processing [12-17]. The second falls to developmental cognitive neuropsychology, and is answered by models of concept learning [18-20]. Such models must assume, as noted above, that the foundational concept of a distinct, manipulable three-dimensional object that persists through time cannot be inferred from observations alone and hence must be encoded by developmental mechanisms acting in earliest infancy; identification of the precise time and mechanism by which this concept emerges is an active area of research [21-24]. These neurofunctional, neurocomputational and cognitive models rely on experimentally-accessible nonlinear mechanisms sensitive to small details in both the genetics and the interaction histories of individual observers. They thus provide a basis for understanding the fundamental constraint that decompositions are not arbitrary.

Zurek [2] distinguishes observers from "the rest of the universe" and in particular from apparatus on the basis of their being "*aware* of the content of their memory" later glossed by saying that "(T)heir information processing machinery ... can readily consult the content of their memory" (p. 759, italics in original). The above considerations suggest that observers are also characterized by distinct memory contents: in particular, observers must remember, and their computational systems must access, von Neumann decompositions. An observer ready state shares with an apparatus ready state the conditions of being turned on, being properly oriented, and having sufficient memory and processing resources; but while the history of an apparatus, e.g. the history of its construction, provides it with a particular information-processing structure, the history of an observer must provide it both with an information-processing structure and with re-identifiable and re-usable representations that are required components of its ready state. Any apparatus reports results; observers are required to report results *and* the decompositions from which they were inferred.

## 5. Three questions of interpretation

The concept of a consistent von Neumann decomposition developed above can now be applied to questions of interpretation. Let us begin with the *gedankenexperiment* proposed by Zeh [25], in which "objects carrying some primitive form of 'core consciousness'" are passed through a double-slit apparatus (p. 232). Zeh clearly intends these "objects" to be observers; they are to have an "elementary awareness of their path through the slits." Observers must have ready states, ready states encode von Neumann decompositions, and von Neumann decompositions specify the parameters of apparatus and certain characteristics of both the environment and the observer. The decomposition employed by Zeh's observer is simplified in that there is no $S$ of interest, only the double-slit apparatus $A$. The only interaction of interest is that between $O$ and $A$, mediated by $E$; hence there are two Hamiltonians $H_{EA}$ and $H_{EO}$ that share environmental eigenstates. Zeh's observer observes at least sporadically as it passes though the apparatus; hence it is interacting with the environment while "in flight."

From Zeh's observer's own perspective, it clearly cannot pass through both slits, as its von Neumann decomposition requires it to represent the slits (i.e. $A$) as re-identifiable over time by itself (i.e. by $O$), and itself as re-identifiable over time both by itself and by external observers. From our external perspective, both Zeh's observer and the slits are interacting with some environmental medium (i.e. $E$), and that environmental medium must have sufficient degrees of freedom to encode spatial information about the slits at a resolution sufficient to provide Zeh's observer with path information. For path information to be well defined for Zeh's observer, these environmental degrees of freedom must, at minimum, be dense enough to decohere positional states of the slits; our perspective, by imposing an informational requirement at the scale of Zeh's observer, must agree with Zeh's observer's perspective on this point. But the slits are larger than Zeh's observer, so a density of environmental degrees of freedom sufficient to decohere positional states of the slits is also sufficient to decohere positional states of Zeh's observer. Hence our perspective agrees with that of Zeh's observer: neither of us reports that Zeh's observer travelled through both slits.

While Zeh's *gedankenexperiment* explores the consequences of considering **S** to be an observer, Zurek's [1, 2] concept of the environment as a "witness" to observations, and its extension to the notion that the universe **U** itself can be considered an observer that ultimately guarantees a world that appears classical [4], explores the consequences of considering **E** to be an observer. These concepts can be interpreted trivially by noting that the universe contains multiple observers, and other observers form part of **E** from the perspective of any given **O**. Zurek appears, however, to intend the more robust interpretation that **E** – and by extension **U** – functions effectively as an observer independently of any "ordinary" observers embedded in it. This interpretation requires an analysis of the state of **E** under which it can be regarded as encoding a von Neumann decomposition. In particular, it requires an analysis of **E** that reveals a bounded representation of an apparatus.

Consider an experimental context in which a number of instrumentation components are visibly connected by cabling. An observer makes measurements by interacting with the final component of the chain; suppose it is a computer screen. This observer **O** encodes as part of her ready state a decomposition that specifies the complete set of component instruments as the apparatus **A**, and regards the rest of the universe (except for the implicitly specified system **S**) as the environment **E**. Suppose now that **O** finds the results displayed on the computer screen perplexing, and hypothesizes that something is awry in the instrumentation set-up. She proceeds to disconnect some cabling and observe the output from the $N^{th}$ component instrument with an oscilloscope; let **A'** represent the entire instrumentation array up to this $N^{th}$ component. If we treat the oscilloscope as part of **O** (this is no different from treating her glasses as part of **O**), then **O**'s new ready state encodes a new decomposition {**S**,**A'**,**E'**,**O**}, where **E'** = **E** + (**A** – **A'**). Let us consider the difference between **E** and **E'**. First, the number of degrees of freedom of **E'** is negligibly larger than that of **E**; **E** is the rest of the universe. Second, **E'** continues to encode the pointer state of **A**, even though the disconnection of some cabling may have changed the value of the pointer state. Indeed if we consider the state of **E'**\(**A** – **A'**), that is, the state of the degrees of freedom of **E'** that correspond to those of **E**, it is clear that nothing has changed except the encoding of the positions of **O**, the oscilloscope, and some cabling. Hence whereas from **O**'s perspective a boundary has shifting – the boundary of the apparatus – from **E**'s perspective nothing has happened. The *physics* at the **A** – **E** boundary has not changed; it is only **O**'s *representation* of the physics – that is, **O**'s ready state – that has changed.

This *gedankenexperiment* of shifting the boundary of **A** can clearly be repeated arbitrarily, with **O** re-specifying **A'** to be anything in the laboratory that she cares to look at, including anything inside any of the instruments. Each re-specification of **A** changes **O**'s ready state, but none of them alter anything more than the states of previously-encoded positions from the perspective of **E**. In particular, no physical interaction on the originally-specified boundary between **A** and **E** is altered; coherence information and thermal noise continue to leak across this boundary exactly as before. But for **E** to be regarded as representing **A**, some component of **E**'s state would have to change (as some component of **O**'s state must change) every time the boundary changes. Hence **E** cannot be regarded as representing **A**, except in the trivial and implicit sense of saying that **E**'s entire state represents **A** by specifying it as **U\E**. Even this trivial sense is unavailable if one attempts to consider **U** as representing **A**. The concept of the environment as a witness must, therefore, be restricted to the notion that the environment is a witness for observers, that is, for entities that encode von Neumann decompositions as components of their ready states. This notion simply re-states the fact that multiple observers can non-destructive access the same environmental encodings.

With this clarification of the environment's role as a witness, the observer-independent formulation of the measurement problem can be addressed. Bacciagaluppi's [11] statement of the observer-independent formulation in the traditional context of a measurement of electron spin is recent and prominent: "the composite of electron, apparatus and environment will be a sum of a state corresponding to the environment coupling to the apparatus coupling in turn to the value +1/2 for the spin, and of a state corresponding to the environment coupling to the apparatus coupling in turn to the value -1/2 for the spin. So again we cannot imagine what it would mean for the composite system to be described by such a sum" (Sect. 3.1). In this formulation of the measurement problem, it is not the observer's experience that is at issue, but rather the ability of an external theorist to explain "what is meant by" a physical state of affairs.

Note first that the formulation above, were it meant to characterize an observer's perspective on the measurement context, would involve a Case 2 decomposition and would therefore, by the analysis outlined in Sect. 2 above, be inconsistent as a description of measurement. Any observer would characterize the electron's state as decohered by the internal environment of the apparatus, and hence would characterize both the pointer state of the apparatus and its environmental encoding as determinate. Hence this formulation can only be meant to characterize the measurement context as if no observer were present, i.e. independently of any decomposition. The decomposition-independent perspective is, however, the perspective of the environment as discussed above. The environment does not *represent* the electron, the apparatus, itself, or any of the interactions that Bacciagaluppi envisions; the environment only has a state. Without assuming a decomposition, it is not even possible to say that this state is a non-trivial superposition, for doing so involves the imposition of a basis in which the state is not pure. From a consistent decomposition-independent perspective, the only thing that can be said is that the environment – or the universe as a whole – is in some state. This is not the statement of a *problem*; it is merely the statement of a fact.

There is, therefore, no consistent observer-independent formulation of the measurement problem. There only appears to be an observer-independent formulation of the measurement problem if one neglects the implicit decompositions built into the words "electron", "apparatus", and "environment". These words denote systems, and as Zurek [1] notes, without systems "the problem of measurement cannot even be stated" (p. 1818). In an environment that is simply regarded as full of interacting stuff, the measurement problem cannot even be stated.

## 6. Conclusions

This paper has considered some consequences of viewing decompositions, and in particular von Neumann decompositions, not as arbitrary extra-theoretical assumptions but as physical states that encode the information that an observer must possess to perform, report and repeat measurements. The primary consequence of this physical view of decompositions is that a system can only be self-consistently represented as a functioning apparatus, with the re-identifiability and information-encoding characteristics that such a representation implies, if it is represented as containing an internal environment that decoheres the states of any of its components that are directly coupled to a quantum system of interest in the course of the measurement. This constraint forces a hard choice: a system can be consistently represented either as linearly coupled to a quantum system (i.e. as being in a quantum superposition) or as functioning as an apparatus with which an observer can repeatably interact, but not both. The traditional von Neumann chain that transfers quantum coherence linearly from the system through the apparatus into the environment and eventually to the observer is not a self-consistent representation of measurement. The measurement problem cannot, on pain of inconsistency, arise from the perspective of either the observer or the environment. The measurement problem is an artifact of an inconsistent description of a physical object as both linearly coupled to a quantum system and as functioning as an apparatus.

The analysis put forward in this paper does not question or challenge the validity of minimal no-collapse quantum mechanics in any way. It does not interpret quantum mechanics. It merely imposes a constraint: decompositions describing measurement interactions must represent apparatus as containing internal decohering environments. All practical descriptions of apparatus, quantum computers, and cognitive systems of observers already satisfy this constraint. It is only descriptions employed pedagogically and in foundational and philosophical discussions that fail to satisfy it. The analysis put forward in this paper demotivates, by deconstructing the measurement problem, interpretive constructs such as Everett branches or extra-quantum mechanical collapse mechanisms, and arguments that quantum mechanics somehow requires special causal roles for consciousness or free will. On the other hand, it motivates the view that measurement can be fully described purely in terms of known physics and unproblematic assumptions concerning the interaction histories of observers. It proposes to begin this description with a characterization of von Neumann decompositions as experimentally-accessible components of the ready states of observers.

From a philosophical perspective, what the present analysis points out is that distinctions must be made between a chunk of the universe, that chunk of the universe bounded by the use of a name, noun, mathematical symbol, or description, and that chunk of the universe named or indicated in a way that presumes or assigns a function or other characteristic. Names impose boundaries, and boundaries implicitly specify remainders or "environments" that are *not* what is under discussion, and which therefore do not have to be attended to in detail. Not attending to something in detail is the information-theoretic formulation of decoherence: coherence information is "lost" only because we have agreed to ignore it when we agree to a name and hence to the boundary of an environment. Failure to draw these distinctions leads to failure to notice that merely bounding an electron as a particle forces it to be described as having a position, bounding the interacting stuff in a rectangular region of space as a visually re-identifiable thing forces it to be described as a macroscopic object, and calling such a bounded thing an apparatus forces it to be described as having a repeatably informative function. But this point about language does not have anything to do with quantum mechanics; it has been discussed at least since Heraclitus. Quantum mechanics is just particularly vulnerable to the unfortunate consequences of not bearing these distinctions in mind.

**References**

1. Zurek, W.: Decoherence, einselection, and the existential interpretation (the rough guide). Phil. Trans. Royal Soc. London A356, 1793-1821 (1998)

2. Zurek, W.: Decoherence, einselection, and the quantum origins of the classical. Rev. Mod. Phys. 75, 715-775. arXiv:quant-ph/0105127v3 (2003)

3. Tegmark, M. Importance of quantum decoherence in brain processes. Phys. Rev. E61, 4194-4206. arXiv:quant-ph/9907009v2 (2000)

4. Zurek, W.: Decoherence and the transition from quantum to classical – revised. Los Alamos Sci. 27, 2-25 (2002)

5. Schlosshauer, M: Decoherence, the measurement problem, and interpretations of quantum mechanics. Rev. Mod. Phys. 76, 1267-1305. arXiv:quant-ph/0312059v4 (2004)

6. Schlosshauer, M: Decoherence and the Quantum to Classical Transition. Berlin: Springer (2007)

7. Hagan, S., Hameroff, S., Tuszynski, J.: Quantum computation in brain microtubules? Decoherence and biological feasibility. Phys. Rev. E65, 861-901 (2002)

8. Hameroff, S: The brain is both neurocomputer and quantum computer. Cog. Sci. 31, 1035-1045 (2007)

9. Schlosshauer, M.: Experimental motivation and empirical consistency in minimal no-collapse quantum mechanics. Ann. Phys. 321, 112-149. arXiv:quant-ph/0506199v3 (2006)

10. Wooters, W., Zurek, W.: A single quantum cannot be cloned. Nature 299, 802-803 (1982)

11.  Bacciagaluppi, G.: The role of decoherence in quantum mechanics.  SEP (2007)

12. Robsertson, L. C.: Binding, spatial attention, and perceptual awareness.  Nat. Rev. Neurosci. 4, 93-102 (2003)

13. Courtney, S.: Attention and cognitive control as emergent properties of information representation in working memory.  Cogn. Affect. Behav. Neurosci. 4, 501-516 (2004)

14. Hommel, B.: Event files: Feature binding in and across perception and action.  Trends Cogn. Sci. 8, 494-500 (2004)

15.  Grush, R.: The emulation theory of representation: Motor control, imagery, and perception.  Behav. Brain Sci. 27, 377-442 (2004)

16. Zmigrod, S., Spape, M., Hommel, B.: Intermodal event files: Integrating features across vision, audition, taction, and action.  Psych. Res.

17.  Barsalou, L. W.: Grounded cognition.  Annu. Rev. Psychol. 59, 617-645 (2008)

18.  Gopnik, A., Schulz, L.: Mechanisms of theory formation in young children.  Trends Cog. Sci. 8, 371-377 (2004)

19.  Casey, B., Tottenham, N., Liston, C., Durston, S.: Imaging the developing brain: What have we learned about cognitive development?  Trends Neurosci. 9, 104-110 (2005)

20.  Tenenbaum, J., Griffiths, T., Kemp, C.: Theory-based Bayesian models of inductive learning and reasoning.  Trends Cog. Sci. 10, 309-318 (2006)

21.  Rethinking infant knowledge: Toward an adaptive process account of successes and failures in object permanence.  Psych. Rev. 104, 686-713 (1997)

22.  Bremer, J. G.: Developmental relationships between perception and action in infancy.  Infant Behav. Devel. 23, 567-582 (2000)

23.  Moore, M. K, Meltzoff, A. N.: Factors affecting infants' manual search for occluded objects and the genesis of object permanence.  Infant Devel. Behav. 31, 168-180 (2008)

24.  Meary, D., Kitromilides, E., Mazens, K., Graff, C., Gentaz, E.: Four-day-old human neonates look longer at non-biological motions of a single point of light.  PloS ONE 2, e186.  doi:10.1371/journal.pone.0000186 (2008)

25.  Zeh, H. D.: The problem of conscious observation in quantum mechanical description.  Found. Phys. Lett. 13, 221-233.  arXiv:quant-ph/9908084v3 (2000)